\documentclass{ws-ijmpe}

\usepackage[normalem]{ulem} % \sout{old text} for strikeout
\usepackage[dvips]{color} % For blue in-text comments and additions
\renewcommand\sout{\bgroup \color{red} \ULdepth=-.5ex \ULset}

\begin{document}

\markboth{LIE-WEN CHEN, CHE MING KO, BAO-AN LI, and GAO-CHAN
YONG}{Determining the density dependence of the nuclear symmetry
energy using heavy-ion reactions }

\catchline{}{}{}{}{}
%%%%%%%%%%%%%%%%%%%%%%%%%%%%%%%%%%%%%%%%%%%%%%%%%%%%%%%%%%%%%%%%%%%%

\title{DETERMINING THE DENSITY DEPENDENCE OF THE NUCLEAR SYMMETRY ENERGY
USING HEAVY-ION REACTIONS}

\author{LIE-WEN CHEN}

\address{Institute of Theoretical Physics, Shanghai Jiao Tong University, \\
Shanghai 200240, China\\
lwchen@sjtu.edu.cn}

\author{CHE MING KO}

\address{Cyclotron Institute and Physics Department, Texas A\&M University, \\
College Station, Texas 77843-3366, USA\\
ko@comp.tamu.edu}

\author{BAO-AN LI}

\address{Department of Physics, Texas A\&M University-Commerce, \\
Commerce, Texas 75429, USA \\
Bao-An\_Li@TAMU-Commerce.edu}

\author{GAO-CHAN YONG}

\address{Institute of Modern Physics, Chinese Academy of
Science, Lanzhou 730000, China\\
yonggaochan@impcas.ac.cn}

\maketitle

\begin{history}
\received{(received date)}
\revised{(revised date)}
%\accepted{(Day Month Year)}
%\comby{(xxxxxxxxxx)}
\end{history}

\begin{abstract}
We review recent progress in the determination of the
subsaturation density behavior of the nuclear symmetry energy from
heavy-ion collisions as well as the theoretical progress in
probing the high density behavior of the symmetry energy in
heavy-ion reactions induced by high energy radioactive beams. We
further discuss the implications of these results for the nuclear
effective interactions and the neutron skin thickness of heavy
nuclei.
\end{abstract}

\section{Introduction}

One of the most important properties of a nuclear matter is its
equation of state (EOS). For symmetric nuclear matter with equal
numbers of protons and neutrons, its EOS has been relatively well
determined from the study of the nuclear giant monopole resonances
(GMR) \cite{You99} as well as the measurements of collective flows
\cite{pawel02} and subthreshold kaon production \cite{Fuc06a} in
nucleus-nucleus collisions. On the other hand, our knowledge on
the EOS of the isospin asymmetric nuclear matter with unequal
numbers of protons and neutrons, especially the part related to
the nuclear symmetry energy, is still largely uncertain
\cite{pawel02,ireview98,ibook,bom,diep03,lat04,baran05,steiner05}.
Although the nuclear symmetry energy at normal nuclear matter
density is known to be around $30$ \textrm{MeV} from the empirical
liquid-drop mass formula \cite{myers,pomorski}, its values at
other densities, particularly at supra-normal densities, are
poorly known \cite{ireview98,ibook}. Advances in radioactive
nuclear beam facilities provide, however, the possibility to pin
down the density dependence of the nuclear symmetry energy in
heavy ion collisions
induced by these nuclei \cite{ireview98,ibook,baran05,li97,fra1,fra2,xu00,tan01,bar02,betty,lis,li00,li02,chen03,chen03b,ono03,liu03,chen04,li04a,shi03,li04prc,rizzo04,Gai04,liyong05,lizx05a,lizx05b,LiQF05b,ma04,tian05,traut06,li06plb,Yon06a,Yon06b,LiBA07a}%
.

In the present talk, we review recent progress in extracting the
information on the subsaturation density behavior of the nuclear
symmetry energy from heavy-ion collisions, especially from the
analysis of isospin diffusion data \cite{tsang04,chen05,li05}.
Furthermore, we discuss the implications the constrained symmetry
energy has for the nuclear effective interactions and the neutron
skin thickness of heavy nuclei. In addition, we also review the
theoretical progress on probing the high density behavior of the
symmetry energy in heavy-ion reactions induced by high energy
radioactive beams.

\section{The nuclear symmetry energy}

The EOS of isospin asymmetric nuclear matter, given by its binding
energy per nucleon, can be generally written as
\begin{equation}
E(\rho ,\alpha )=E(\rho ,\alpha =0)+E_{\mathrm{sym}}(\rho )\alpha
^{2}+O(\alpha ^{4}),  \label{EsymPara}
\end{equation}%
where $\rho =\rho _{n}+\rho _{p}$ is the baryon density with $\rho
_{n}$ and $\rho _{p}$ denoting the neutron and proton densities,
respectively; $\alpha =(\rho _{n}-\rho _{p})/(\rho _{p}+\rho
_{n})$ is the isospin asymmetry; $E(\rho ,\alpha =0)$ is the
binding energy per nucleon in symmetric nuclear matter, and
\begin{equation}
E_{\mathrm{sym}}(\rho )=\frac{1}{2}\frac{\partial ^{2}E(\rho
,\alpha )}{\partial \alpha ^{2}}\vert_{\alpha =0}  \label{Esym}
\end{equation}
is the nuclear symmetry energy. Neglecting contributions from
higher-order terms in Eq. (\ref{EsymPara}) leads to the well-known
empirical parabolic law for the EOS of asymmetric nuclear matter.
For nuclear matter at moderate densities, this parabolic law has
been verified in results from all many-body theories. As a good
approximation, the symmetry energy can thus be extracted from the
binding energy difference between pure neutron matter and
symmetric nuclear matter, i.e., $E_{\mathrm{sym}}(\rho )\approx
E(\rho ,\alpha =1)-E(\rho ,\alpha =0)$. It should be mentioned
that possible presence of higher-order terms in $ \alpha$ at
supra-normal densities could significantly modify the proton
fraction in $\beta$-equilibrium neutron-star matter and thus the
critical density for the direct Urca process that is responsible
for faster cooling of neutron stars \cite{Zha01,Ste06}.

Around the nuclear matter saturation density $\rho _{0}$, the
nuclear symmetry energy $E_{\mathrm{sym}}(\rho )$\ can be expanded
to second-order in density as
\begin{equation}
E_{\mathrm{sym}}(\rho )=E_{\mathrm{sym}}(\rho
_{0})+\frac{L}{3}\left( \frac{\rho -\rho _{0}}{\rho _{0}}\right)
+\frac{K_{\mathrm{sym}}}{18}\left( \frac{\rho -\rho _{0}}{\rho
_{0}}\right) ^{2},  \label{EsymLK}
\end{equation}
where $L$ and $K_{\mathrm{sym}}$ are the slope and curvature
parameters of the nuclear symmetry energy at $\rho _{0}$, i.e.,
\begin{equation}
L=3\rho _{0}\frac{\partial E_{\text{sym}}(\rho )}{\partial \rho
}|_{\rho
=\rho _{0}},\text{ }K_{\text{sym}}=9\rho _{0}^{2}\frac{\partial ^{2}E_{\text{%
sym}}(\rho )}{\partial ^{2}\rho }|_{\rho =\rho _{0}}.
\label{LKsym}
\end{equation}%
The $L$ and $K_{\mathrm{sym}}$ characterize the density dependence
of the nuclear symmetry energy around normal nuclear matter
density, and thus carry important information on the properties of
nuclear symmetry energy at both high and low densities. In
particular, the slope parameter $L$ has been found to correlate
linearly with the neutron-skin thickness of heavy nuclei and thus
can in principle be determined from measured thickness of the
neutron skin of such nuclei
\cite{diep03,brown00,hor01,typel01,furn02,kara02,steiner05b,chen05nskin}.
Unfortunately, because of the large uncertainties in the
experimental measurements, no reliable information on the slope
parameter of the nuclear symmetry energy has so far been obtained
in this way.

At the saturation density and around $\alpha =0$, the isobaric
incompressibility of asymmetric nuclear matter can also be
expressed to second-order in $\alpha $ as \cite{Pra85,Lop88}
\begin{equation}
K(\alpha )\approx K_{0}+K_{\mathrm{asy}}\alpha ^{2},  \label{Kasy}
\end{equation}
where $K_{0}$ is the incompressibility of symmetric nuclear matter
at the saturation density and the isospin-dependent part
$K_{\mathrm{asy}}\approx K_{\mathrm{sym}}-6L $ \cite{bar02}
characterizes the density dependence of the nuclear symmetry energy.
Information on $K_{\mathrm{asy}}$ can in principle be extracted
experimentally by measuring the GMR in neutron-rich nuclei.

\section{Constraining the subsaturation density behavior of the
symmetry energy from isospin diffusion data}

\begin{figure}[htb]
\begin{minipage}{13.5pc}
\includegraphics[scale=0.62]{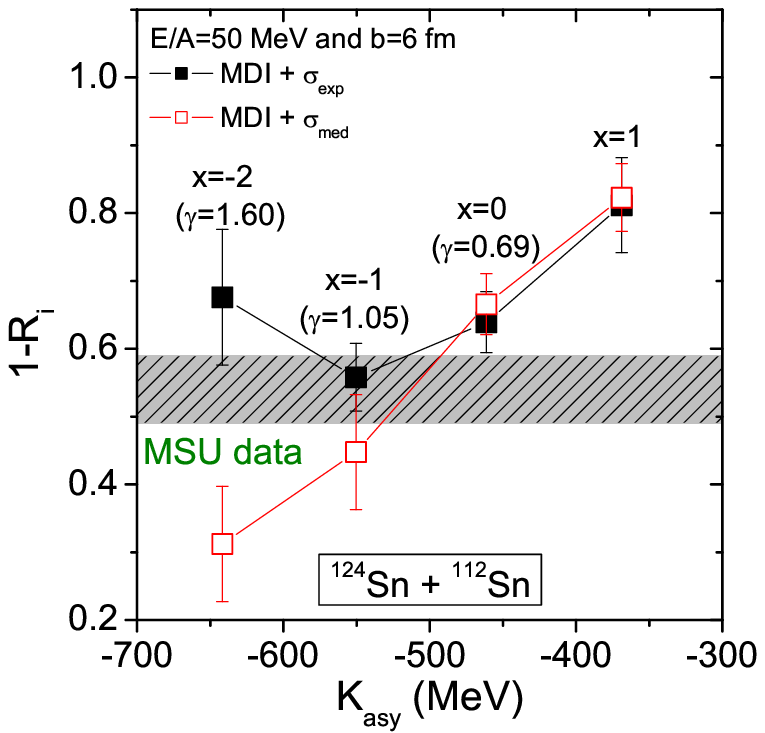}
\end{minipage}
\begin{minipage}{13.5pc}
\includegraphics[scale=0.62]{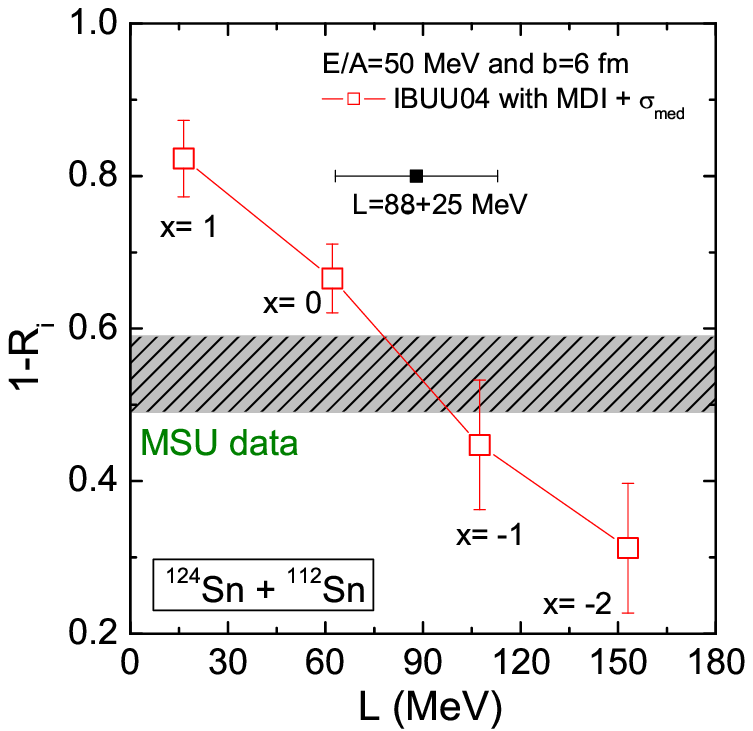}
\end{minipage}
\caption{(Color online) Degree of the isospin diffusion $1-R_{i}$ as
functions of $K_{\mathrm{asy}}$ (left panel, taken from Ref.
\protect\cite{li05}) and $L$ (right panel, taken from Ref.
\protect\cite{chen05nskin}) using the MDI interaction with $x=-2$,
$-1$, $0$, and $1$.} \label{RiKasyL}
\end{figure}

The symmetry energy is known to affect the isospin diffusion in
intermediate-energy heavy ion collisions \cite{shi03}.
Experimentally, the degree of isospin diffusion between the
projectile nucleus $A$ and the target nucleus $B$ can be studied
via \cite{tsang04,rami}
\begin{equation}
R_{i}=\frac{2X^{A+B}-X^{A+A}-X^{B+B}}{X^{A+A}-X^{B+B}}, \label{Ri}
\end{equation}%
where $X$ is any isospin-sensitive observable. By construction,
the value of $R_{i}$ is $1~(-1)$ for symmetric $A+A~(B+B)$
reaction. For asymmetric reactions, the value of $R_i$ decreases
from $1$ to about zero as the colliding nuclei approaches isospin
equilibrium. In the NSCL/MSU experiments with $A=$ $^{124}$Sn and
$B=$ $^{112}$Sn at a beam energy of $50$ MeV/nucleon and an impact
parameter about $6$ fm, the isospin asymmetry of the
projectile-like residue was used as the isospin tracer $X$
\cite{tsang04}. Using an isospin- and momentum-dependent IBUU04
transport model with experimental free-space or in-medium
nucleon-nucleon (NN) cross sections, the dependence of $R_{i}$ on
the nuclear symmetry energy in these reactions was studied in
Refs. \cite{chen05,li05} for different isospin- and
momentum-dependent MDI interactions \cite{das03}. Shown in Fig.
\ref{RiKasyL} are the results for the degree of the isospin
diffusion $1-R_{i}$ as functions of $K_{\mathrm{asy}}$ (left
panel) and $L$ (right panel). The shaded band in Fig.
\ref{RiKasyL} indicates the data from NSCL/MSU \cite{tsang04}. For
the experimental free-space NN cross sections, comparing
theoretical results with experimental data allows us to extract a
nuclear symmetry energy of $E_{\text{sym}}(\rho )\approx 31.6(\rho
/\rho _{0})^{1.05}$, corresponding to the parameter $x=-1$ in the
MDI interaction. With the medium-dependent NN cross sections,
which are important for isospin-dependent observables
\cite{li05,li05a}, to explain the isospin diffusion data requires
an even softer nuclear symmetry energy of $E_{\text{sym}}(\rho
)\approx 31.6(\rho /\rho _{0})^{\gamma }$ with $\gamma \approx
0.69 - 1.05$ \cite{li05}. In this case, the extracted value for
the slope parameter of the nuclear symmetry energy at saturation
density is $L=88\pm 25$ MeV and that for the isospin-dependent
part of the isobaric incompressibility of isospin asymmetric
nuclear matter is of $K_{\mathrm{asy}}=-500\pm 50$ MeV
\cite{chen05,li05,chen05nskin}.

The above extracted symmetry energy agrees with the symmetry
energy $E_{\mathrm{sym}}(\rho )=31.6(\rho /\rho _{0})^{0.69}${,
corresponding to $L\approx 65$ MeV and $K_{\mathrm{asy}}\approx
-453$ MeV, that was} recently obtained from the isoscaling
analyses of isotope ratios in intermediate energy heavy ion
collisions \cite{She07}. The extracted value of
$K_{\mathrm{asy}}=-500\pm 50$ MeV from the isospin diffusion data
is also consistent with the value $K_{\mathrm{asy}}=-550\pm 100$
MeV obtained from recently measured isotopic dependence of the GMR
in even-A Sn isotopes \cite{Gar07}. These empirically extracted
values for $L$ and $K_{\rm sym}$ represent the best current
constraints on the nuclear symmetry energy at sub-normal
densities.

\section{Constraining effective interactions in the SHF and RMF models}

The non-relativistic Skyrme-Hartree-Fock (SHF)
\cite{brown00,brack85,fried86,brown98,clw99,stone03} and
relativistic mean-field (RMF) \cite{Ser86,Rei89,Rin96,Ser97,Men06}
models are two extensively used phenomenological approaches for
nuclear structure studies \cite{Ben03}. In these models, a number
of parameters are adjusted to fit the properties of nuclear matter
as well as those of many finite nuclei. Since there are many
different parameter sets in the SHF and RMF models, it is of
interest to see to what extend these parameter sets are consistent
with the constrained symmetry energy from heavy-ion collisions.

\begin{figure}[htb]
\begin{minipage}{13.5pc}
\includegraphics[scale=0.62]{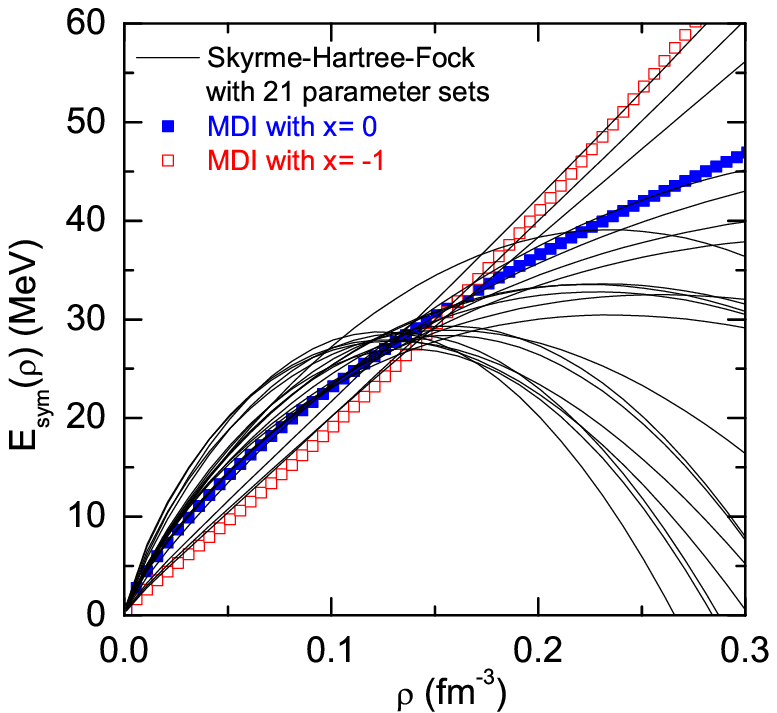}
\end{minipage}
\hspace*{10pt}
\begin{minipage}{13.5pc}
\includegraphics[scale=0.62]{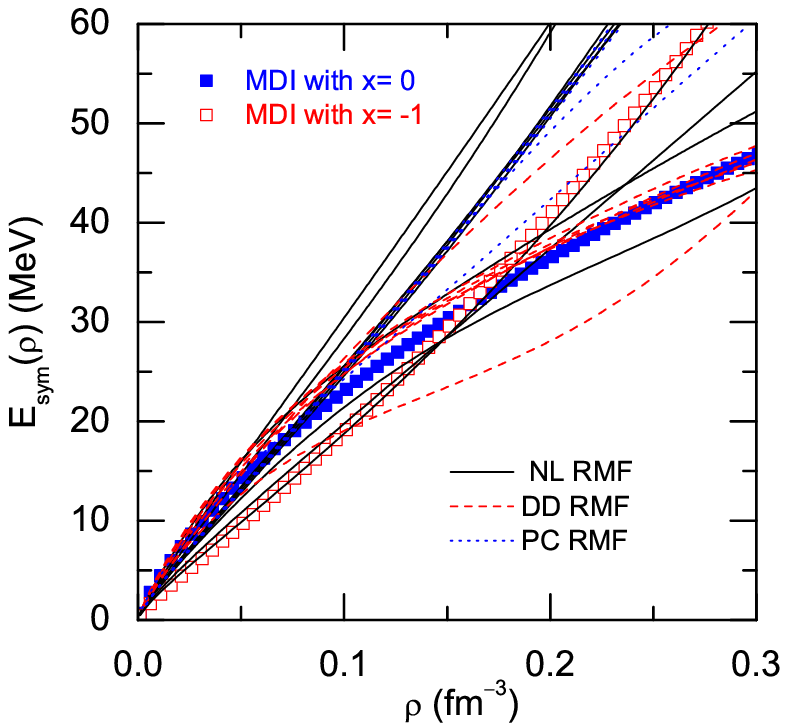}
\end{minipage}
\caption{\protect\small (Color online) Density dependence of
$E_{\text{sym}}(\rho )$ for $21$ sets of Skyrme interaction
parameters (left panel, taken from Ref.
\protect\cite{chen05nskin}) and for $23$ parameter sets in the
nonlinear, density-dependent, and point-coupling RMF models (right
panel, taken from Ref. \protect\cite{Che07}). } \label{EsymSHFRMF}
\end{figure}

Fig. \ref{EsymSHFRMF} displays the density dependence of
$E_{\text{sym}}(\rho )$ for $21$ sets of Skyrme interaction
parameters \cite{chen05nskin}, i.e., \textrm{SKM},
\textrm{SKM}$^{\ast }$, $\mathrm{RATP}$, \textrm{SI},
\textrm{SII}, $\mathrm{SIII}$, \textrm{SIV}, \textrm{SV},
\textrm{SVI}, \textrm{E}, \textrm{E}$_{\sigma }$,
\textrm{G}$_{\sigma }$, $\mathrm{R}_{\sigma }$, $\mathrm{Z}$,
$\mathrm{Z}_{\sigma }$, $\mathrm{Z}_{\sigma }^{\ast }$,
$\mathrm{T}$, $\mathrm{T3}$, $\mathrm{SkX}$, $\mathrm{SkXce}$, and
$\mathrm{SkXm}$ (left panel), as well as for $23$ parameter sets
in the nonlinear (i.e., NL1 \cite{Lee86}, NL2 \cite{Lee86}, NL3
\cite{Lal97}, NL-SH \cite{Sha93}, TM1 \cite{Sug94}, PK1
\cite{Lon04}, FSU-Gold \cite{Tod05}, HA \cite{Bun03}, NL$\rho $
\cite{Liu02}, NL$\rho \delta $ \cite{Liu02}), density-dependent
(i.e., TW99 \cite{Typ99}, DD-ME1 \cite{Nik02}, DD-ME2
\cite{Lal05}, PKDD \cite{Lon04}, DD \cite{Typ05}, DD-F
\cite{Kla06}, and DDRH-corr \cite{Hof01}), and point-coupling
(i.e., PC-F1 \cite{Bur02}, PC-F2 \cite{Bur02}, PC-F3 \cite{Bur02},
PC-F4 \cite{Bur02}, PC-LA \cite{Bur02}, and FKVW \cite{Fin06}) RMF
models \cite{Che07} (right panel). Values of the parameters in the
Skyrme interactions can be found in Refs.
\cite{brack85,fried86,brown98}. For comparison, we also show in
Fig. \ref{EsymSHFRMF} results from the MDI interactions with
$x=-1$ (open squares) and $0$ (solid squares) which give,
respectively, the upper and lower bounds for the stiffness of the
nuclear symmetry energy from the isospin diffusion data
\cite{li05}. It is seen that the density dependence of the
symmetry energy varies drastically among different interactions
and most of the parameter sets give symmetry energies that are
inconsistent with the extracted one except SIV, SV, G$_\sigma$,
and R$_\sigma$ among the Skyrme parameters \cite{chen05nskin} and
TM1, NL$\rho $, NL$\rho \delta $, PKDD, and FKVW among the RMF
parameter sets \cite{Che07}.

\section{Predictions on the neutron skin thickness of heavy
nuclei}

Also affected by the density dependence of nuclear symmetry energy
is the neutron skin thickness $S$ of a nucleus, which is defined
as the difference between the root-mean-square radii of neutron
distribution and proton distribution. In particular, the skin
thickness of a nucleus is sensitive to the slope parameter $L$ of
the nuclear symmetry energy at normal nuclear matter density
\cite{diep03,brown00,hor01,typel01,furn02,kara02,chen05}. Using
above $21$ sets of Skyrme interaction parameters, the neutron skin
thickness of several nuclei has been studied and it is found that
for heavy nuclei there exists a strong linear correlation between
$S$ and $L$. For $^{208}$Pb, the linear correlation between $S$
and $L$ is given by the following expression \cite{chen05}:
\begin{equation}
S(^{\text{208}}\text{Pb)}=(0.1066\pm 0.0019)+(0.00133\pm
3.76\times 10^{-5})\times L,  \label{SLPb208a}
\end{equation}%
\begin{equation}
L=(-78.5\pm 3.2)+(740.4\pm 20.9)\times S(^{\text{208}}\text{Pb)},
\label{SLPb208b}
\end{equation}%
where the units of $L$ and $S$ are \textrm{MeV} and \textrm{fm},
respectively. For $^{132}$Sn and $^{124}$Sn, the corresponding
expressions are \cite{chen05}
\begin{equation}
S(^{\text{132}}\text{Sn)}=(0.1694\pm 0.0025)+(0.0014\pm 5.12\times
10^{-5})\times L,  \label{SLSn132a}
\end{equation}%
\begin{equation}
L=(-117.1\pm 5.4)+(695.1\pm 25.3)\times S(^{\text{132}}\text{Sn)},
\label{SLSn132b}
\end{equation}%
and
\begin{equation}
S(^{\text{124}}\text{Sn)}=(0.1255\pm 0.0020)+(0.0011\pm 4.05\times
10^{-5})\times L,  \label{SLSn124a}
\end{equation}%
\begin{equation}
L=(-110.1\pm 5.2)+(882.6\pm 32.3)\times S(^{\text{124}}\text{Sn)},
\label{SLSn124b}
\end{equation}%
Similar linear relations between $S$ and $L$ are also expected for
other heavy nuclei. This is not surprising since the thickness of
the neutron skin in heavy nuclei is determined by the pressure
difference between neutron and proton matters in a nucleus, which
is proportional to the parameter $L$
\cite{diep03,brown00,hor01,typel01,furn02,kara02,chen05}.

The extracted $L$ value from the isospin diffusion data in heavy ion
collisions allows us to determine from Eqs. (\ref{SLPb208a}%
), (\ref{SLSn132a}), and (\ref{SLSn124a}), respectively, a neutron
skin thickness of $0.22\pm 0.04$ fm for $^{208}$Pb, $0.29\pm 0.04$
fm for $^{132}$Sn, and $0.22\pm 0.04$ fm for $^{124}$Sn. We note
these results are not only in surprisingly good agreement with
those obtained from the RMF model using an accurately calibrated
relativistic parametrization, which can describe simultaneously
the ground state properties of finite nuclei and their monopole
and dipole resonances \cite{Tod05}, but also consistent with the
experimental data \cite{sta94,clark03,kras04}.

\section{Probing the high density behavior of the nuclear symmetry energy in
heavy-ion reactions}

\label{highdensity}

Although significant progress has been made in the determination
of the density dependence of the nuclear symmetry energy at
sub-normal densities, the high density behavior of the the nuclear
symmetry energy is still poorly known. Fortunately, heavy-ion
reactions, especially those induced by high energy radioactive
beams to be available at high energy radioactive beam facilities,
provide a unique opportunity to pin down the high density behavior
of the symmetry energy. In this section, we illustrate via
transport model simulations several experimental observables which
are sensitive to the high density behavior of the symmetry energy.

Most observables proposed so far for studying the density
dependence of the nuclear symmetry energy employ differences or
ratios of isospin multiplets of baryons and mesons as well as
mirror nuclei, such as the neutron/proton ratio of emitted
nucleons \cite{li97}, neutron-proton differential flow
\cite{li00}, neutron-proton correlation function \cite{chen03},
$t$/$^{3}$He \cite{chen03b,lizx05a}, $\pi ^{-}/\pi ^{+}$
\cite{li02,Gai04,liyong05,LiQF05b}, $\Sigma ^{-}/\Sigma ^{+}$
\cite{lizx05b} and $K^{0}/K^{+}$ ratios
\cite{LiQF05c,Fer05,Fer06}, etc. However, some of these
observables, especially those involving neutrons, pose great
challenges in experimental measurements. The measurement of
neutrons, particularly the low energy ones, always suffers from
low detection efficiencies even for the most advanced neutron
detectors. Therefore, observables involving neutrons normally have
large systematic errors. Moreover, for essentially all these
observables, the Coulomb force on charged particles plays an
important role and sometimes competes strongly with the symmetry
potential. It is thus very desirable to find experimental
observables which can reduce the influence of both the Coulomb
force and the systematic errors associated with the measurement of
neutrons. A possible candidate for such an observable is the
double ratios of emitted particles taken from two reaction systems
using four isotopes of same element, such as the neutron/proton
ratio in the neutron-rich system over that in the more symmetric
system, as recently done by Famiano {\it et al.} \cite{Lyn06}.
Theoretically, the systematic errors associated with transport
model calculations, which are mostly related to the uncertainties
in the in-medium NN cross sections, techniques of treating
collisions, sizes of the lattices in calculating the phase space
distributions, techniques in handling the Pauli blocking, etc.,
are also expected to be reduced for relative observables that
involve double ratios from two similar reaction systems.

\begin{figure}[htb]
\begin{minipage}{13.5pc}
\includegraphics[width=6cm,height=3.5cm]{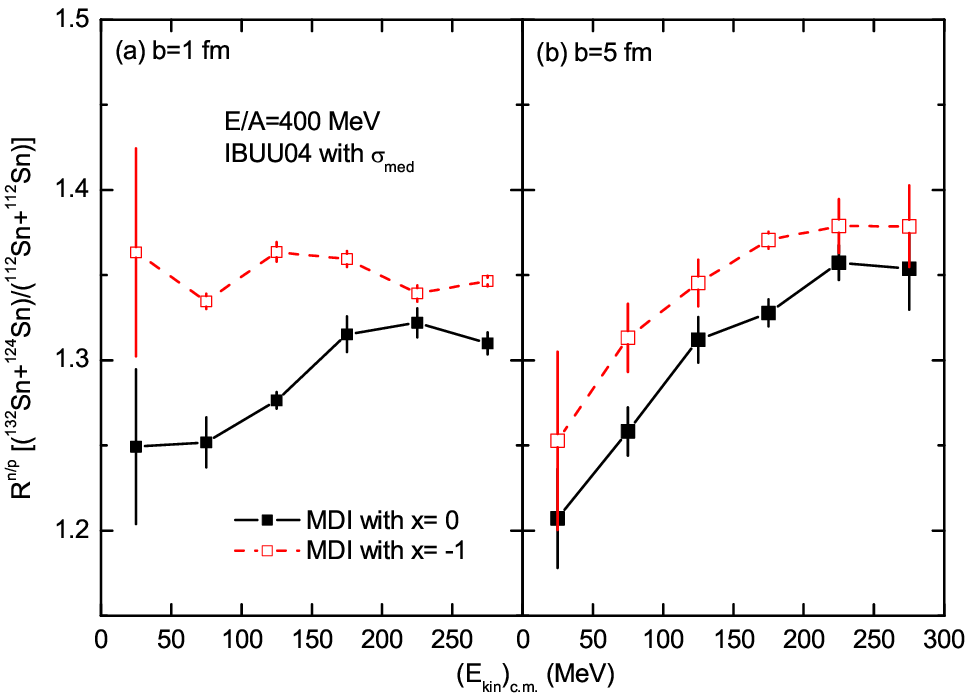}
\end{minipage}
\hspace*{10pt}
\begin{minipage}{13.5pc}
\includegraphics[width=6cm,height=4.2cm]{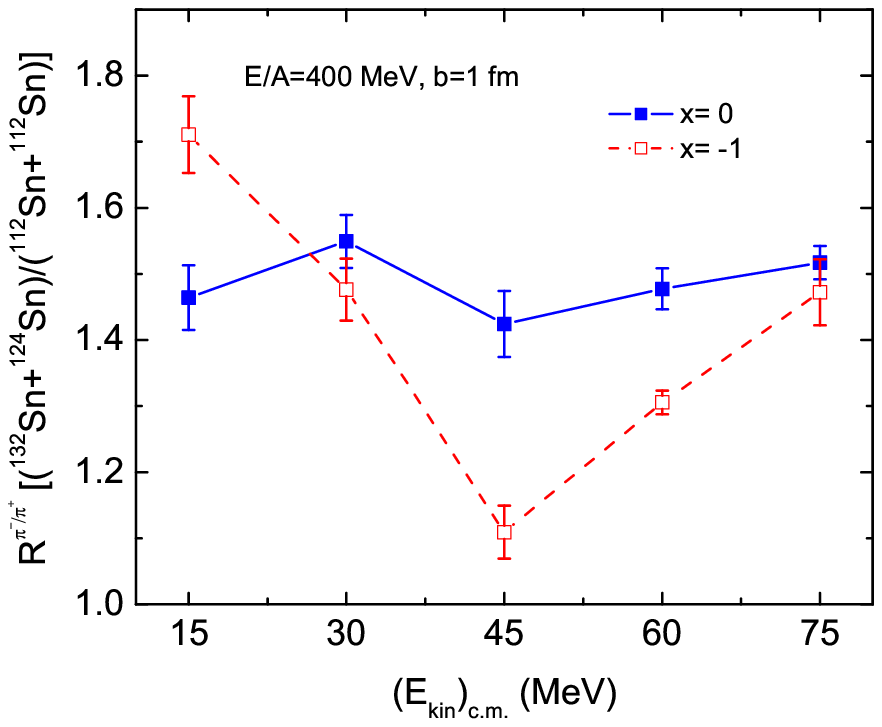}
\end{minipage}
\caption{(Color online) Double ratio for neutron/proton (left
panels, taken from Ref. \protect\cite{li06plb}) and $\pi ^{-}/\pi
^{+}$ (right panel, taken from Ref. \protect\cite{Yon06a}) in
$^{132}$Sn$+^{124}$Sn and $^{112}$Sn$+^{112}$Sn reactions at $400$
MeV/nucleon.} \label{DRnppion}
\end{figure}

\begin{figure}[htb]
\begin{minipage}{13.5pc}
\includegraphics[width=6cm,height=4.2cm]{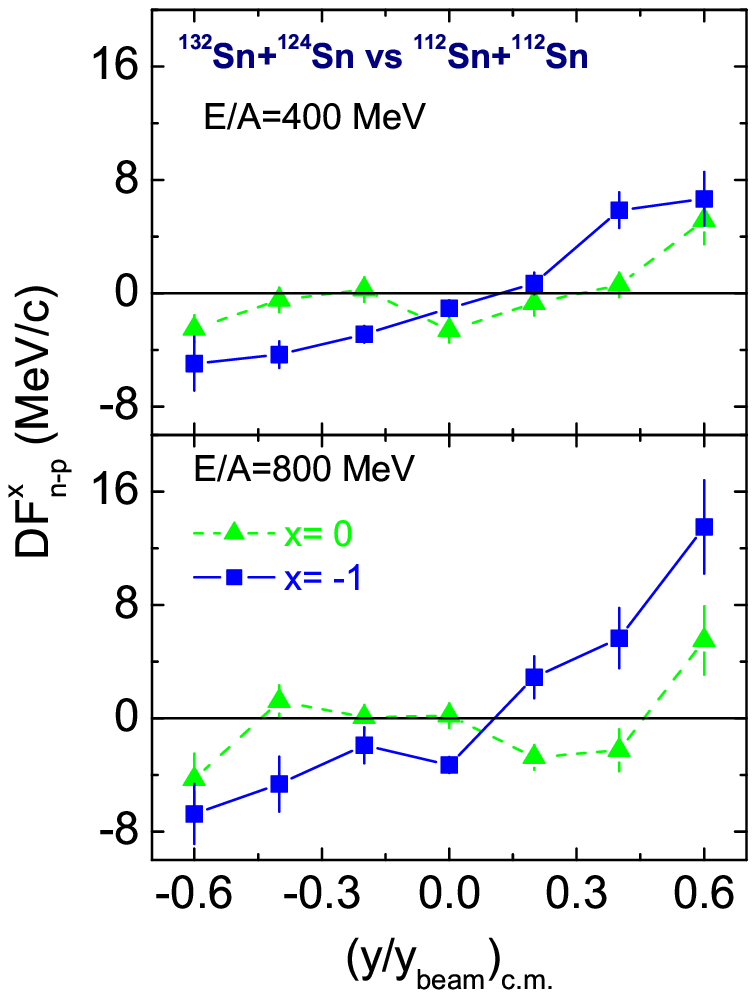}
\end{minipage}
\hspace*{5pt}
\begin{minipage}{13.5pc}
\includegraphics[width=6cm,height=4.2cm]{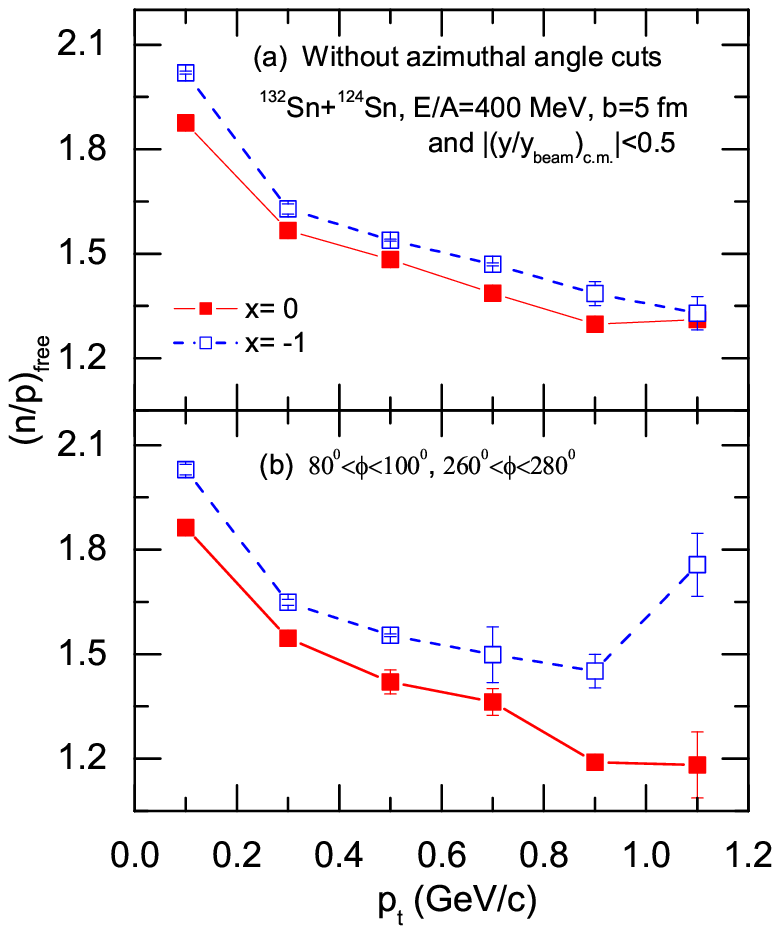}
\end{minipage}
\caption{(Color online) Left panels: Rapidity dependence of the
double n-p differential transverse flow in reactions of Sn+Sn
isotopes at $400$ and $800$ MeV/nucleon with $b=5$ fm (Taken from
Ref. \protect\cite{Yon06b}). Right panels: Transverse momentum
dependence of the ratio of midrapidity neutrons to protons emitted
in the reaction of $^{132}$Sn+$^{124} $Sn at $400$ MeV/nucleon and
$b=5$ fm, with (lower window) and without (upper window) an
azimuthal angle cut (Taken from Ref. \protect\cite{Yon07}).}
\label{DdiffFnpRnpSq}
\end{figure}

Both the double neutron/proton ratio \cite{li06plb} and the double
$\pi ^{-}/\pi ^{+}$ ratio \cite{Yon06a} in $^{132}$Sn$+^{124}$Sn
and $^{112}$Sn$+^{112}$Sn reactions at $400$ MeV/nucleon have been
studied recently in the IBUU04 model in order to demonstrate the
effect of symmetry energy at high density, and the results are
shown in Fig.\ \ref{DRnppion}. It is seen that these ratios have
about the same sensitivity to the density dependence of symmetry
energy as the corresponding single ratio in the respective
neutron-rich system. Also, the double neutron-proton differential
flow \cite{Yon06b} in $^{132}$Sn$+^{124}$Sn and
$^{112}$Sn$+^{112}$Sn reactions and the n/p ratio of the
squeeze-out nucleons \cite{Yon07} in $^{132}$Sn$+^{124}$Sn have
also recently been studied in the IBUU04 model. As shown in Fig.\
\ref{DdiffFnpRnpSq}, both are indeed sensitive to the symmetry
energy. In particular, compared to other potential probes, the n/p
ratio of squeeze-out nucleons carries more direct information
about the symmetry energy at high densities. The sensitivity to
the high density behavior of the nuclear symmetry energy observed
in the n/p ratio of squeeze-out nucleons with high transverse
momenta is probably the highest found so far among all observables
studied within the same transport model.

\begin{figure}[htb]
\begin{minipage}{13.5pc}
\includegraphics[width=5.5cm,height=3.8cm]{RKaonFe.eps}
\end{minipage}
\hspace*{20pt}
\begin{minipage}{13.5pc}
\vspace{-0.3cm}
\includegraphics[width=6cm,height=4.3cm]{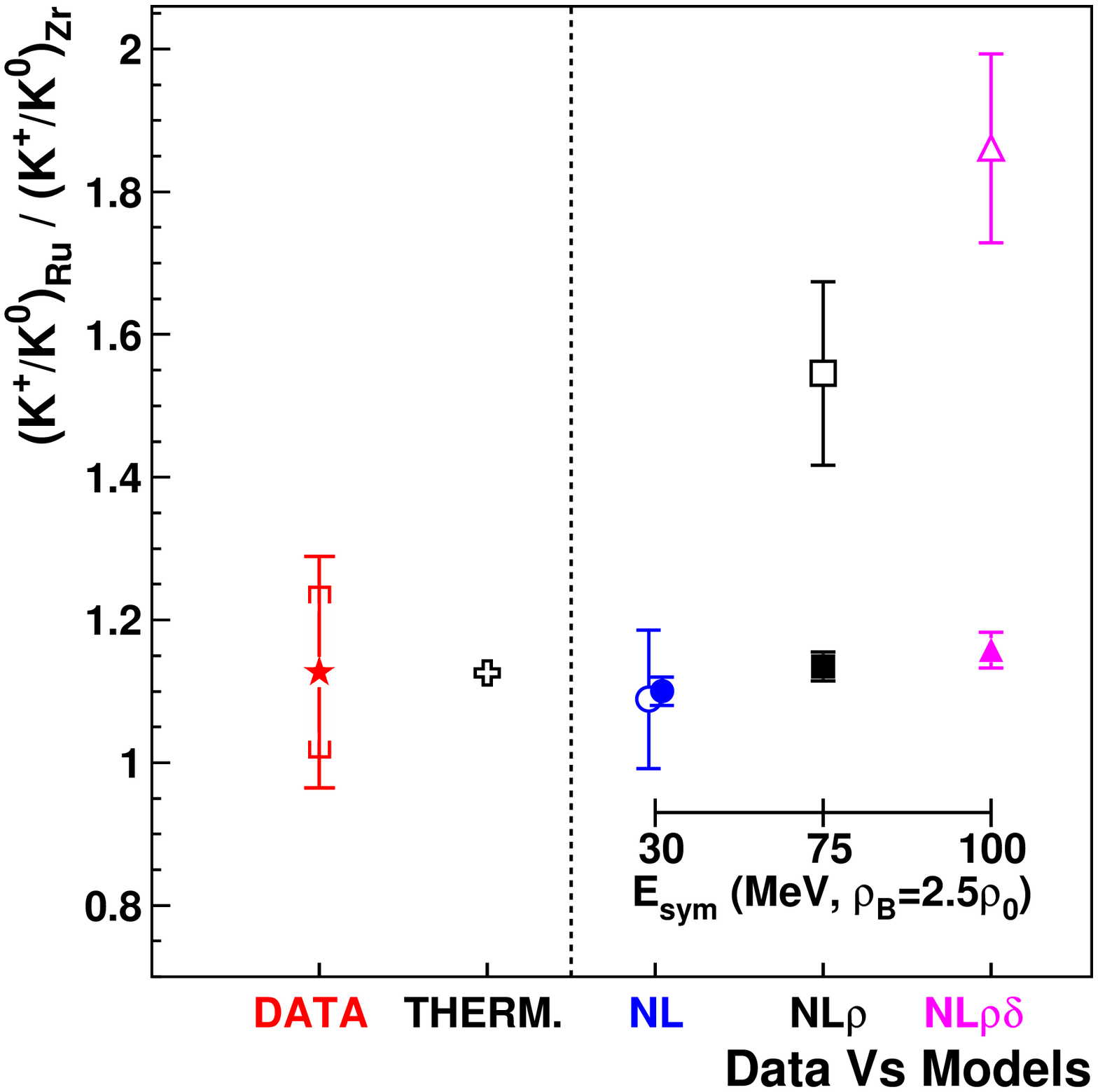}
\end{minipage}
\caption{(Color online) Left panel: $\protect\pi ^{-}/\protect\pi
^{+}$ (upper) and $K^{+}/K^{0}$ (lower) ratios as a function of
the incident energy for central Au+Au collisions with the RBUU
model (Taken from Ref. \protect\cite{Fer05}). Right panel:
Experimental ratio ($K^{+}/K^{0}$)$_{Ru}$/($K^{+}/K^{0}$)$_{Zr}$
(star) and theoretical predictions of the thermal model (cross)
and the transport model with nuclear matter (open symbols) and
realistic (filled symbols) collision scenarios (Taken from Ref.
\protect\cite{Lop07}).} \label{RKaon}
\end{figure}

Since the proposal of Aichelin and Ko \cite{Aic85} that the kaon
yield in heavy ion collisions at energies that are below the
threshold for kaon production in a nucleon-nucleon collision in
free space may be a sensitive probe of the EOS of nuclear matter
at high densities, a lot of works have been done both
theoretically and experimentally on this problem
\cite{Fuc06a,Cas90,Ko96,Ko97,Cas99,Kol05}. As the kaon is an
iso-doublet meson with the quark content of $d\overline{s}$ for
$K^{0}$ and $u\overline{s}$ for $K^{+}$, the $K^{0}/K^{+}$ ratio
provides a potentially good probe of the nuclear symmetry energy
as the $n/p$ and $\pi^-/\pi^+$ ratios. The $K^{0}/K^{+}$ ratio is
expected to be particularly sensitive to the high density behavior
of nuclear symmetry energy as kaons are produced mainly from the
high density region during the early stage of the reaction and
thus suffer negligible absorption effects. The symmetry energy
effect on the $K^{0}/K^{+}$ ratio in heavy-ion collisions has
recently been investigated using the UrQMD \cite{LiQF05c} and RBUU
models \cite{Fer05,Fer06}. The results from the RBUU model
\cite{Fer05}, shown in the left panel of Fig.\ \ref{RKaon},
indicate that at beam energies below and around the kinematical
threshold of kaon production, the $K^{0}/K^{+}$ inclusive yield
ratio is more sensitive to the symmetry energy than the $\pi
^{-}/\pi ^{+}$. Subthreshold kaon production thus could provide a
promising tool to extract information on the density dependence of
the nuclear symmetry energy.

Experimentally, the FOPI collaboration at GSI-Darmstadt has
recently reported the results on $K^{+}$ and $K^{0}$ meson
production in $_{44}^{96}$Ru + $_{44}^{96}$Ru and $_{40}^{96}$Zr +
$_{40}^{96}$Zr collisions at a beam kinetic energy of $1.528$ $A$
GeV \cite{Lop07}. The measured double ratio
($K^{+}/K^{0}$)$_{Ru}$/($K^{+}/K^{0}$)$_{Zr}$ is compared in the
right panel of Fig.\ \ref{RKaon} to the predictions from both the
thermal model and the RBUU transport model using two different
collision scenarios and under different assumptions on the
stiffness of the symmetry energy. The experimental data show a
good agreement with the thermal model prediction and also with
that of the transport model using the collision scenario of an
infinite nuclear matter with a soft symmetry energy. Although more
realistic transport simulations of the collisions show a similar
agreement with the data, they exhibit a rather weak sensitivity to
the symmetry energy. Due to the complexity of subthreshold kaon
production in heavy-ion collisions \cite{Kol05,Fuc06a}, further
experimental and theoretical studies are needed to extract useful
information on the high density behavior of the nuclear symmetry
energy from subthreshold kaon production in heavy-ion collisions
induced by neutron-rich nuclei.

\section{Summary}

Heavy-ion reactions induced by neutron-rich nuclei provide a
unique means to investigate the equation of state of
isospin-asymmetric nuclear matter, especially the density
dependence of the nuclear symmetry energy. We have reviewed the
recent progress in extracting the information on the subsaturation
density behavior of the nuclear symmetry energy from heavy-ion
collisions. From the recent analysis of the isospin diffusion data
in heavy-ion collisions using an isospin- and momentum-dependent
transport model with in-medium NN cross sections, a value of
$L=88\pm 25$ MeV for the slope parameter of the nuclear symmetry
energy at saturation density and a value of
$K_{\mathrm{asy}}=-500\pm 50$ MeV for the isospin-dependent part
of the isobaric incompressibility of isospin asymmetric nuclear
matter have been extracted. The extracted symmetry energy agrees
with the symmetry energy $E_{\mathrm{sym}}(\rho )=31.6(\rho /\rho
_{0})^{0.69}$, corresponding to $L\approx 65$ MeV and
$K_{\mathrm{asy}}\approx -453$ MeV, that was recently obtained
from the isoscaling analyses of isotope ratios in intermediate
energy heavy ion collisions \cite{She07}. The extracted value of
$K_{\mathrm{asy}}=-500\pm 50$ MeV from the isospin diffusion data
is also consistent with the value $K_{\mathrm{asy}}=-550\pm 100$
MeV obtained from recently measured isotopic dependence of the GMR
in even-A Sn isotopes \cite{Gar07}. These empirically extracted
values for $L$ and $K_{\rm sym}$ represent the best current
constraints on the nuclear symmetry energy at sub-normal
densities. This in turn imposes strong constraints on the
parameters in both Skyrme and RMF effective interactions. Detailed
studies have indicated that only SIV, SV, G$_\sigma$, and
R$_\sigma$ among the Skyrme parameters and only TM1, NL$\rho $,
NL$\rho \delta $, PKDD, and FKVW among the RMF parameter sets have
symmetry energies that are consistent with the extracted one.
Furthermore, the extracted $L$ value from the isospin diffusion
data has led to predicted neutron skin thickness of $0.22\pm 0.04$
fm for $^{208}$Pb, $0.29\pm 0.04$ fm for $^{132}$Sn, and $0.22\pm
0.04$ fm for $^{124}$Sn.

We have also reviewed recent theoretical progress in identifying
the observables in heavy-ion collisions induced by high energy
radioactive nuclei that are sensitive to the high density behavior
of the symmetry energy. Many potentially sensitive probes have
been found, and they include the $\pi ^{-}/\pi ^{+}$ ratio,
isospin fractionation, n-p differential flow, double n/p and $\pi
^{-}/\pi ^{+}$ ratios, double n-p differential transverse flow as
well as the $K^{0}/K^{+}$ and $\Sigma ^{-}/\Sigma ^{+}$ ratios.
Studying these observable in future experiments at high energy
radioactive beam facilities is expected to provide significant
constraints on the behavior of the symmetry energy at supra-normal
densities.

\section*{Acknowledgments}
This work was supported in part by the National Natural Science
Foundation of China under Grant Nos. 10575071 and 10675082, MOE of
China under project NCET-05-0392, Shanghai Rising-Star Program
under Grant No. 06QA14024, the SRF for ROCS, SEM of China, the
China Major State Basic Research Development Program under
Contract No. 2007CB815004, the US National Science Foundation
under Grant Nos. PHY-0457265 and PHY-0652548, the Welch Foundation
under Grant No. A-1358, and the Research Corporation under Award
No. 7123.

\end{document}